\documentclass{aastex}
\usepackage{emulateapj5}
\usepackage{natbib}
\usepackage{amsmath}

\newcommand{\sbf}{\ensuremath{\sigma_{\rm bf}}}
\newcommand{\keV}{\ensuremath{\rm keV}}

\newcommand{\nH}{\ensuremath{N_{\rm H}}}
\newcommand{\cm}{\ensuremath{\rm cm}}
\newcommand{\avg}[1]{\ensuremath{\langle #1 \rangle}}
\newcommand{\sigism}{\ensuremath{\sigma_{\rm ISM}}}
\newcommand{\Nbar}{\ensuremath{\avg{N}}}
\newcommand{\sigbar}{\ensuremath{\avg{\sigma}}}
\newcommand{\dd}{\ensuremath{{\rm d}}}

\begin{document}

\received{}
\accepted{}

\slugcomment{Astrophysical Journal, in press}
\shorttitle{X-Ray Absorption in the ISM}
\shortauthors{Wilms, Allen, \& McCray}

\title{On the Absorption of X-rays in the Interstellar Medium}

\author{J.~Wilms}
\affil{Institut f\"ur Astronomie und Astrophysik,
  Abt.~Astronomie}
\affil{Waldh\"auser Stra\ss{}e 64, D-72076 T\"ubingen, Germany}
\email{wilms@astro.uni-tuebingen.de} 
\and
\author{A.~Allen, R.~McCray}
\affil{JILA, University of
  Colorado}
\affil{Campus Box 440, Boulder, CO~80309-0440, USA}
\email{allenu@super.colorado.edu, dick@jila.colorado.edu}

\begin{abstract}
  We present an improved model for the absorption of X-rays in the
  interstellar medium (ISM) intended for use with data from future
  X-ray missions with larger effective areas and increased energy
  resolution such as \textsl{Chandra} and the X-ray Multiple Mirror
  mission, in the energy range $\gtrsim 100$\,eV.  Compared to
  previous work, our formalism includes recent updates to the
  photoionization cross section and revised abundances of the
  interstellar medium, as well as a treatment of interstellar grains
  and the H$_2$ molecule. We review the theoretical and observational
  motivations behind these updates and provide a subroutine for the
  X-ray spectral analysis program XSPEC that incorporates our model.
\end{abstract}


\section{Introduction}
The precise knowledge of the modification of X-ray radiation in the
interstellar medium (ISM) is of crucial importance for the
understanding of X-ray spectra from cosmic sources since the observed
X-ray spectra have to be corrected for this modification before
interpreting the observed data.  The current state of the art in the
computation of the energy-dependent photoionization cross section of
the ISM, $\sigism$, has been summarized by \citet[BM92]{balu:92a}.
These authors greatly improved\footnote{When referring to the cross
  sections of BM92 we use their subroutines as implemented as the
  \texttt{phabs} model in XSPEC, version 10.0z \citep{arnaud:96a}.
  This implementation contains improvements compared to the original
  publication. Starting with XSPEC, version 11.0, other cross-sections
  are used in the computation of this model (see~\S\ref{sec:atomic}
  below).}  the precision in the computation of $\sigism$ compared to
previous implementations
\citep[MM83]{strom:61a,brown:70a,fireman:74a,ride:77a,morrison:83a} by
allowing to adjust the assumed abundances of the ISM and by using
polynomial fits to the photoionization cross sections of
\citet{henke:82a}.

Although the BM92 model provides sufficient precision for the data
analysis with current instruments, the advent of X-ray missions with
instruments of high energy resolution and large effective areas, such
as \textsl{Chandra} or  the Newton X-ray Multiple Mirror mission
(XMM-Newton), will result in large improvements in the precision of X-ray
astronomical measurements.  It is therefore necessary to have tools at
hand that reflect the current understanding of the physics of the
interaction of X-rays with the ISM at the percent level.

In this paper we improve on the models of BM92 by incorporating
recent improvements in the photoionization cross sections and by
taking into account the physics of the interstellar dust and the
molecular phase of the interstellar medium.
Section~\ref{sec:model} describes the formalism employed to
compute $\sigism$. We describe the atomic physics data
(\S\ref{sec:atomic}) and explain our choices of elemental and
molecular abundances (\S\ref{sec:abundances}) as well as dust grain
parameters (\S\ref{sec:grain}).  We summarize the model, compare
it to previous models, and point out its remaining shortcomings
in Section~\ref{sec:summary}.  In the appendices, we derive the
influence of dust on $\sigism$ (Appendix~\ref{sec:f}) and
describe the implementation of our model into XSPEC
(Appendix~\ref{sec:xspec}).

The intended readership of this paper consists of users of the
new missions, many of whom do not intend to become specialists on
the ISM. We have therefore attempted to include sufficient
information to make this paper self-contained, and have chosen a
review-like style.

\section{Modeling the X-ray Absorption in the ISM}\label{sec:model}

\subsection{X-ray Absorptivity in the ISM}\label{sec:int}
The total photoionization cross section of the ISM, $\sigism$, is
obtained by summing over the contributions of the astrophysically
relevant elements. Taking into account the phases of the ISM,
$\sigism$ can be written as
\begin{equation}\label{eq:sigism}
\sigism = \sigma_{\rm gas} + \sigma_{\rm
  molecules} + \sigma_{\rm grains}. 
\end{equation}
As is common for X-ray applications, we normalize $\sigism$ to the
total hydrogen number density (i.e., in molecular, neutral, or ionized
form), $N_{\rm H}$, so that the observed X-ray spectrum of a source,
$I_{\rm obs}$, is given by
\begin{equation}
I_{\rm obs}(E) = {\rm e}^{-\sigism(E)\,N_{\rm H}}\,I_{\rm
    source}(E),
\end{equation}
where $N_{\rm H}$ is measured in atoms\,cm$^{-2}$ and $I_{\rm source}(E)$ is
the X-ray spectrum as emitted by the X-ray source. We drop
the explicit energy dependence of $\sigism$ for the rest of the paper.

The contribution of the gaseous, non-molecular cold and warm
phases of the ISM to $\sigism$ is obtained by summing the
photoionization cross sections of the individual atoms and ions
in these phases, weighting their contributions by the abundances.
Hence,
\begin{equation}
\sigma_{\rm gas} =  \sum_{Z,i} A_Z \cdot a_{Z,i} \cdot
(1-\beta_{Z,i}) \cdot \sbf(Z,i) 
\end{equation}
where we use the notation pioneered by \citet{ride:77a}, in which 
$A_Z=N(Z)/N({\rm H})$ is the abundance in number of element $Z$
with respect to hydrogen, $a_{Z,i}=N(Z,i)/N(Z)$ is the fraction
of ions of element $Z$ that are in ionization stage $i$, and
$\sbf(Z,i)$ is the total photoionization cross section of element
$Z$ in ionization stage $i$. The depletion of the elements into
grains is taken into account through the ``depletion factor''
$1-\beta_{Z,i}$.

For the molecular phase of the ISM, only molecular hydrogen needs
to be taken into account due to its large abundance ($\sim 20\%$
of the hydrogen might be molecular, see \S\ref{sec:abundances}).
Therefore,
\begin{equation}
\sigma_{\rm molecules}= A_{\rm H_2} \sbf({\rm H_2}).
\end{equation}
For material in the grain phase of the ISM, several complications
arise.  Since the X-ray optical depth of a large grain can be
much greater than one (see Fig.~\ref{fig:graintau}), most of the
X-ray absorption will occur on its surface.  Consequently, the
contribution of material ``within'' the grain to $\sigism$ is
small.  Thus, the net effect of this
shielding is that the presence of grains \emph{reduces} the
effective absorptivity of the ISM relative to the absorptivity of
a completely gaseous medium.  As we show in Appendix~\ref{sec:f},
the total optical depth for grains with a size distribution $\dd
n_{\rm gr}(a)/\dd a$ is given by
\begin{equation}\label{eq:taudist}
\begin{split}
\tau_{\rm grains} &=\sigma_{\rm grains} N_{\rm H} \\
    &=N_{\rm H} \xi_{\rm g}  
  \int_0^\infty \frac{\dd n_{\rm gr}(a)}{\dd a}  \sigma_{\rm geom} 
  \left(1-\exp\left(-\sigbar \Nbar\right)\right)
\dd a
\end{split}
\end{equation}
In the derivation of equation~\eqref{eq:taudist} we assume that the
grains are of chemically homogeneous composition with an average
photoabsorption cross section $\sigbar$, and can be approximated as
spheres with radius $a$ and average column density $\Nbar$ (measured
in atoms\,cm$^{-2}$). Since the grains are partly transparent to
X-rays, their absorption cross section is smaller than the grain
geometrical cross section, $\sigma_{\rm geom}=\pi a^2$.  Finally, the
number of grains per hydrogen atom along the line of sight is
$\xi_{\rm g}$. We discuss the validity of these assumptions in
Section~\ref{sec:grain} and give explicit formulae for all parameters
of equation~\eqref{eq:taudist} in Appendix~\ref{sec:f}.

In the following sections we describe the data used for evaluating
Eqs.~\eqref{eq:sigism} and~\eqref{eq:taudist} in greater detail.

\subsection{Atomic Physics}\label{sec:atomic}
Despite decades of effort, the knowledge of elemental
photoionization cross sections in the X-ray regime is still 
uncertain. In many cases, therefore, one has to rely on either
theoretical work or interpolation along the isoelectronic
sequence to obtain usable cross sections, and then use
compilations of experimental data to gauge the precision of these
theoretical computations. In what follows we will describe the basis
of the photoionization cross sections used here, focusing on the 
region $E \gtrsim 100$\,eV. The emphasis is on
the choice of cross sections that are good for instruments with a
resolution comparable to the CCDs on \textsl{Chandra} or
XMM-Newton. Therefore, we will not make an attempt to include the
resonances that are important close to the absorption edges that might
well be important for the grating instruments aboard these
satellites. 

For the computation of the photoionization cross section of H, we
use the fitting formula of \citet{band:90a}, which provides a
very good approximation of the exact, analytic cross section but
is much faster to evaluate. The maximum deviation between the
\citet{band:90a} formula and the analytical cross section
\citep[e.g.,][eq.~71.7, divided by two]{bethe:57a} is $\sim
1.5\%$ at energies below 2\,keV.

The cross section for helium is taken from the recent theoretical
evaluation of \citet{yan:98a}.  Autoionization resonances
occur at energies less than 0.08\,keV, and the four strongest 
resonances are included here (as in BM92),
using values from \citet{oza:86a} and \citet{fernley:87a}.
For energies between 0.1\,keV and 1\,keV, the helium cross
section is \emph{smaller} by a factor of $\sim 0.9$ to $\sim 0.6$
than that assumed in the subroutine used by BM92, which is based
on theoretical work of \citet{chen:91b} and on experimental data
by \citet{marr:76a}.  The \citet{yan:98a} cross section, however,
agrees with other more recent theoretical computations
\citep[e.g., those of][]{verner:95a} and is in agreement with
more recent experimental data \citep[][and references
therein]{yan:98a}.  It also has the correct $E^{-3.5}$ asymptote
required from non-relativistic quantum theory \citep{bethe:57a}.
For implications of this smaller He cross section on studies of
the ISM, see, e.g., \citet{arabadjis:99a}.

For all other elements we use the photoionization cross sections
as originally published by \citet{verner:93a} and later updated
by \citet{verner:95a}, and which are used in XSPEC version 11.0. 
These authors fitted the results from
Hartree-Dirac-Slater computations of $\sbf$ for individual
subshells to a semi-empirical fitting formula. The major
advantage of this semi-empirical fitting formula over previous
fitting formulae (such as, e.g., polynomial fits) is that it
requires only a small number of fitting parameters to express
$\sbf$ to high accuracy while preserving the correct asymptotical
properties.  A comparison between the experimental data and these
cross sections is presented by \citet{verner:95a}, who focus on
the overall agreement throughout the electromagnetic spectrum. In
addition, we compared the total photoionization cross sections
computed from the fitting formula with the compilation of data
presented by \citet{henke:93a} for the range from 0.1\,keV to
10\,keV.  Except for energies close to the K- and L-shell edges,
our comparisons indicate maximum relative deviations between
these cross sections on the order of 5\% above 1\,\keV.

In their determination of $\sbf$ close to the K- and L-shell
edges, \citet{henke:93a} used a linear interpolation through the
resonances to the edge energy. Since this edge energy was
obtained from experimental data, it includes the effects of
energy shifts in the solid state and thus is not representative
of the true edge energy for the gaseous state of the materials.
Indeed, the difference in edge energy between the solid and the
gaseous state can be up to several 10\,eV \citep[see, e.g.,
][for an illustrative experimental example]{nicolosi:91a}.  In
many cases, \citeauthor{henke:93a} also do not consider
the different subshell threshold energies.  On the
other hand, the \citet{verner:95a} cross sections are based on
quantum-mechanical computations that include the resonances, and
the edge energy is unaffected by solid state effects. We have
checked the precision of the \citeauthor{verner:95a} edge
energies by comparing them with the computations of
\citet{gould:91a} which are more appropriate for our work than
the \citet{henke:82a} or \citet{henke:93a} values since they
assume the elements to be in the gas phase.  For $Z>8$ the edge
energies of \citet{verner:95a} agree on the 10\,eV level with
those of \citet{gould:91a}.  For C, N, and O, the edge energies
adopted by \citet{verner:95a} are between those of
\citet{henke:93a} and \citet{gould:91a}. We therefore decided to
adopt the \citeauthor{verner:95a} edge energies for this work.
Note, however, that these differences in the edge energy are
relevant only for instruments with extremely high resolution and
are irrelevant for most other work.

For molecular hydrogen we adopt the cross sections reported by
\citet{yan:98a} for energies above 85\,eV.  These cross sections are
claimed to be accurate to about 5\% and have the correct $E^{-3.5}$
asymptotics. Albeit outside of our formal energy range ($E\gtrsim
100$\,eV) we note that the fit formula given by
\citet[eq.~18]{yan:98a} for energies below 85\,eV has significant
deviations with respect to the tabulated cross sections of
\citet{samson:94a}, on which the \citeauthor{yan:98a} values are
based. Furthermore the fit formula is non-continuous at 85\,eV. 
 For energies between 30 and 85\,eV  we find that the
cross section can be represented by
\begin{equation}\label{eq:h2fit}
\sigma_{{\rm bf},{\rm H}_2}(x)=\sum_{i=0}^5 a_i x^{-i}
\end{equation}
where $x=E/15.4$\,eV and where $\sigma_{\rm bf}$ is given in Mbarn
($=10^{-18}\,\cm^{2}$). The fit coefficients $a_i$ are given in
Table~\ref{tab:h2tab}. The maximum deviation between the fit and the
tabulated data is less than 1\%.  We note that the photoabsorption
cross section of ${\rm H}_2$ is $\sim 2.85\sigma_{\rm H}$ which
results in an increase over previous estimates of $\sigism$ for
regions where molecular hydrogen is an important contributor to the
total absorptivity (i.e., below $\sim$1\,keV).  Molecular effects were
not included in the earlier models, so essentially these models
assumed $\sigma_{{\rm bf}, {\rm H}_2}=2\sigma_{{\rm bf}, {\rm H}}$.

\subsection{Abundances}\label{sec:abundances}
Naturally, the assumed elemental abundances are of crucial importance
for the computation of $\sigism$. While many measurements of gas-phase
abundances have been made, measurement of the total gas-plus-dust
abundance of the ISM is very difficult.  As a result, solar
abundances, i.e., abundances determined from analysis of the solar
photosphere or (carbonaceous) meteorites, have generally been used as
the reference abundance for the ISM \citep[and references
therein]{anders:82a,grevesse:89a,anders:89a,shull:93a}. For reference,
we list in Table~\ref{tab:abund} what we consider to be the best
estimates of these ``local'' ISM abundances.  We list the logarithmic
abundances by number, normalized to a hydrogen abundance of $\log
A_{\rm H}=12$, based on the meteoritic abundances of
\citet{anders:89a}. We updated the abundances of C and N using the
results of \citet{grevesse:91b} and \citet{grevesse:93a},
respectively.  The solar abundance of iron has been the subject of
several discussions in recent years \citep[and references
therein]{raassen:98a,kostik:96a,biemont:91a}. We adopted $\log A_{\rm
  Fe}=7.50\pm 0.05$ as recently determined by \citet{grevesse:99a} in
a critical reevaluation of the available data on \ion{Fe}{1} lines.
The uncertainty of the solar abundances in Table~\ref{tab:abund} is
$\sim0.06$\,dex or smaller with the exception of $\log A_{\rm Ne}$ and
$\log A_{\rm Ar}$, for which the uncertainty is $0.1$\,dex.

As more abundance measurements have been made outside our solar
system, it has become apparent that the total gas plus dust ISM
abundances are actually lower than the solar abundances \citep[and
references therein]{sofia:94a,savage:96b}.  Studies of the carbon
abundance reviewed by \citet{snow:95a} strongly indicate a subsolar
abundance of carbon in the ISM, with a probable value of $\sim 70$\%
solar. Similar reductions of 20\% to 30\% with respect to solar
abundances are also indicated for the other metals
\citep{snow:96a,savage:96b}. This trend is consistent with new
measurements of the gas-phase abundance of O, C, and N obtained with
the Goddard High Resolution Spectrograph (GHRS) on the Hubble Space
Telescope \citep{meyer:98a,meyer:97a,cardelli:96a}.

Possible differences between solar and ISM abundances had been noted
already at the time of publication of previous models for $\sigism$,
and the previous computations of $\sigism$ closely reflect the
historical changes in the understanding of these local abundances.
\citet{ride:77a}, for instance, used abundances that were greater than
solar abundances in order to better represent abundance values in an
ISM assumed to be chemically enriched by galactic evolution, and the
model of BM92 was specifically designed to allow adjustments to the
assumed abundances\footnote{The abundances available in XSPEC are
  those of \citet{anders:89a}, \citet{feldman:92a}, and
  \cite{anders:82a}; in addition, user-specified abundances are
  possible. The BM92 model allows the user to vary the abundances with
  respect to these default abundances.}.  Now that a consistent
picture of lower ISM abundance seems to have emerged, the cosmic
abundance needs to be revised in models of $\sigism$.  To that end we
adopt default recommended abundances for the ISM.  These abundance
values are listed in Table~\ref{tab:abund}, and refer to the
\emph{total} abundance of the elements in either the gaseous or grain
phase of the ISM. For C, N, and O, we use the values of
\citet{cardelli:96a}, \citet{meyer:97a}, and \citet{meyer:98a},
respectively, while for all other elements we use the abundances given
by \citet[Mg, Si, S, Ca, Ti, Cr, Fe, Ni]{snow:96a}, or, where no value
was recommended by these authors, we assume an abundance of 70\% with
respect to our adopted solar abundances.

We are aware that these abundances are still very much uncertain; the
uncertainties in the general ISM abundances are still on the order of
0.1\,dex or higher and also strongly depend on the line of sight.
However, we feel that it has been adequately established that solar
abundances are enriched with respect to the ISM, and that for most
astronomical work using the ISM abundances in Table~\ref{tab:abund} is
preferable. Since one may anticipate the need for further revision of
the interstellar abundances, we have also included an option to change
these default abundances in the XSPEC implementation of our model
(Appendix~\ref{sec:xspec}), as BM92 did. 

Due to the appreciable difference in the X-ray cross sections for
molecular and neutral hydrogen (\S\ref{sec:atomic}), it is
necessary  to include the contribution of molecular
hydrogen.  About half of the hydrogen in the Galaxy as a whole is
molecular; however, it is not uniformly distributed
\citep{shull:82b}.  Surveys of low galactic latitudes indicate
that the ratio between ${\rm H}_2$ and \ion{H}{1} is strongly
dependent on the radial distance from the galactic center
\citep{brinks:90a,bronfman:88a}.  UV observations of H$_2$
absorption lines with \textsl{Copernicus} and with the ORFEUS II
(orbiting and retrievable far- and extreme ultraviolet
spectrometer) echelle spectrograph indicate that about 20\ldots
25\% of the hydrogen is in H$_2$ in the local ISM
\citep{savage:77a,gringel:00a}. These results are very
line-of-sight dependent, though, with some lines of sight being 
completely free of H$_2$. Because of these large fluctuations in
the H$_2$/\ion{H}{1} ratio, it is thus difficult to choose a
``typical value''. In order to include the effect of molecular
hydrogen on $\sigism$ we therefore choose to use the value
of 20\% as the default value.  To more accurately account for the
opacity of molecular hydrogen, however, we strongly recommend
that this ratio be set to that corresponding to the line-of-sight
to the source. The abundances of other molecules are much smaller
so that their inclusion is only necessary for work along line of
sights with very high $N_{\rm H}$ and hence not within the scope
of this paper.

\subsection{Grain Physics}\label{sec:grain}
Evidence for the presence of dust grains in the ISM comes from the
observed interstellar extinction \citep{fitzpatrick:99a,mathis:90a},
as well as from the fact that the observed abundances of the elements
in the ISM are generally less than the abundances assumed for the ISM
\citep{savage:96b}.  The ratio between the observed and the assumed
abundances is called the ``depletion'', $\beta_Z$, of the elements. It
is generally assumed that the depletion indicates the presence of dust
in the ISM, especially since the most depleted elements, C, O, and Si,
are thought to be good candidates for the formation of solids in the
ISM \citep[and references therein]{sofia:94a,snow:95a,mathis:96b}.

We have adopted a simplified grain model which assumes spherical
grains in an MRN size distribution \citep{mathis:77a},
\begin{equation}\label{eq:mrn}
\frac{\dd n_{\rm gr}(a)}{\dd a}\propto a^{-3.5},
\qquad \text{where $a_{\rm min}\leq a \leq a_{\rm max}$},
\end{equation}
where typically grain size ranges from 0.025\,\micron\ to
0.25\,\micron\ \citep{draine:84a}.  While the grain composition and
structure is still a matter of debate
\citep{mathis:96b,dwek:97a,smith:98a,wolff:98a}, here, we assume
composite (``fluffy'') grains, consisting of vacuum inclusions in
solids.  Thus, the grain density in our model ($\rho=1\,\rm
g\,cm^{-3}$) is less than the $\rho=2\,\rm g\,cm^{-3}$ used in
previous models \citep[][, MM83]{fireman:74a,ride:77a}.

We assume chemically homogeneous grains consisting of silicates,
graphites, and oxides.  We list the adopted chemical composition of
the grains in Table~\ref{tab:abund} in terms of the ``depletion
factor'' $1-\beta_Z$, the ratio between the \emph{gas} abundance and
the \emph{total} interstellar abundance of the elements.
Table~\ref{tab:abund} is based on the compilation of measurements
given by \citet{shull:93a}, which is in general agreement with later
GHRS measurements \citep{savage:96b}.  Contrary to
\citeauthor{shull:93a}, we do not assume N to be depleted into grains
as is suggested by the observational evidence
\citep{savage:96b,meyer:97a}. As recommended by \citet{sofia:94a}, we
also assume a larger abundance of Fe in grains than that assumed by
\citet{shull:93a}. For cobalt the values determined by
\citet{mullman:98a} are used.

To illustrate the large uncertainties associated with the grain
composition, we also list in Table~\ref{tab:abund} the depletions used
by \citet{ride:77a} and MM83, the previous works which included grains
in the computation of $\sigism$.  The depletion factors of
\citet{ride:77a}, as shown in Table~\ref{tab:abund}, are based on
gas-abundance measurements using \textsl{Copernicus} data.  Only one
of the depletion factors of MM83 is based on observation, that for
oxygen \citep{boer:79a,boer:81a}; for the rest, MM83 considered
elements to be either completely depleted into grains or entirely in
gas form. This simplifying assumption represents the maximum effect of
grains on the ISM opacity.

We note that grains are included here partly for
completeness and partly to indicate how changes to the current ISM
grain model might affect observations made with the more recent X-ray
observatories, and not as an attempt to model the solid state in the
interstellar medium.  While a ``perfect'' grain model meeting all
abundance and observational requirements remains elusive, our grain
model at least adequately reproduces the observed extinction, emission
features, and depletions of the diffuse ISM \citep{mathis:96b}.  For
more detailed work, the assumption of spherical grains must be
modified \citep{mathis:90a}.

\begin{figure*}
\centerline{\includegraphics[width=0.8\textwidth]{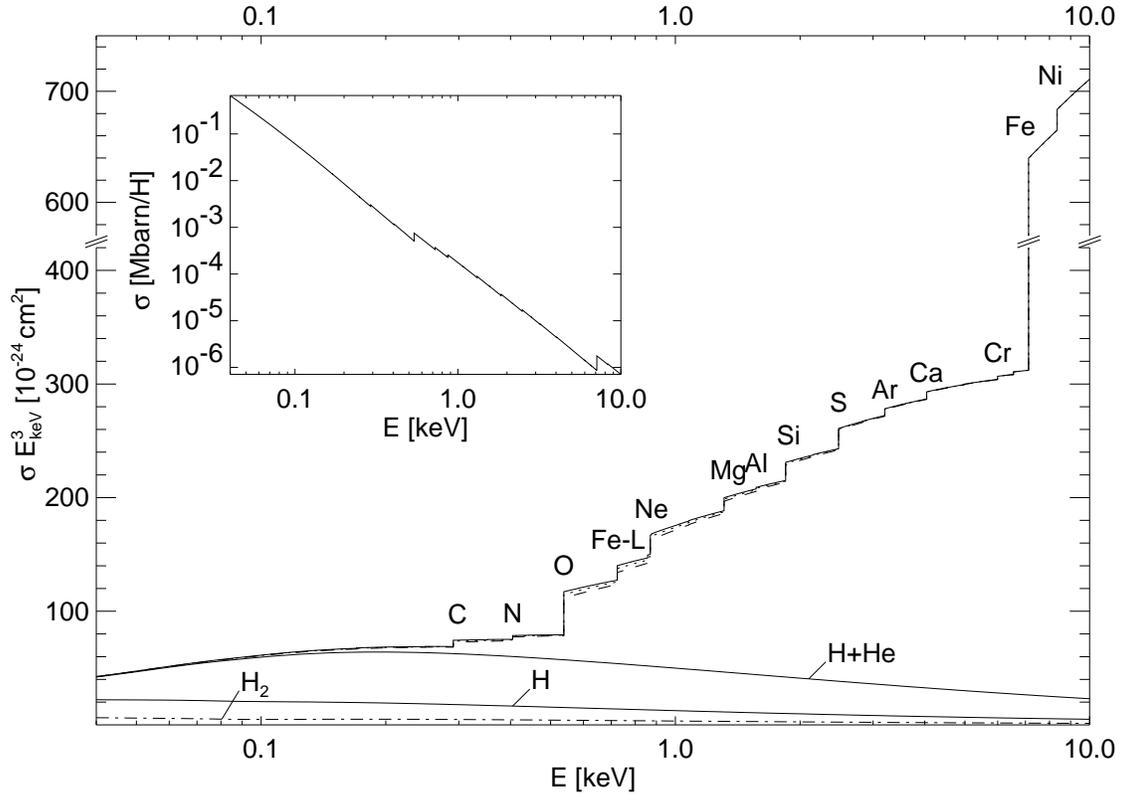}}
\caption{\protect\label{fig:cosab}Absorptivity per hydrogen atom of the ISM using
  the assumptions described in the text. The dotted line is the
  absorptivity including grains with a MRN distribution, and the
  dashed line is the absorptivity assuming that all grains are of
  radius $a=0.3\,\micron$. The inset shows the cross section without
  the multiplication with $E^3$.  We also illustrate the contribution
  of hydrogen and hydrogen plus helium to the total cross section. The
  contribution of the $\rm H_2$ cross section to the total hydrogen
  cross section is indicated by the dot-dashed line.}
\end{figure*}

\section{Conclusions and Summary}\label{sec:summary}
\subsection{Results}
Using equation~\eqref{eq:sigism} and the assumptions described above,
we compute $\sigism$. The results are shown in
Fig.~\ref{fig:cosab}, where we display the X-ray absorptivity as
$\sigism E^3$ to emphasize deviations from the $E^{-3}$
proportionality of $\sbf$.  By reading the plot as a
``bar-diagram'', it is easy to estimate the relative importance
of the contributors to $\sigism$.  For energies above the oxygen
K-edge at $\sim$0.5\,\keV (see Tab.~\ref{tab:abund}), the X-ray
opacity is dominated by the metals, and H and He are relatively
unimportant.  Below 1\,\keV, C, N, O, and Ne are the important
absorbers, while above 1\,\keV, Si, S, and Fe are important.

\begin{figure*}
\centerline{\includegraphics[width=0.8\textwidth]{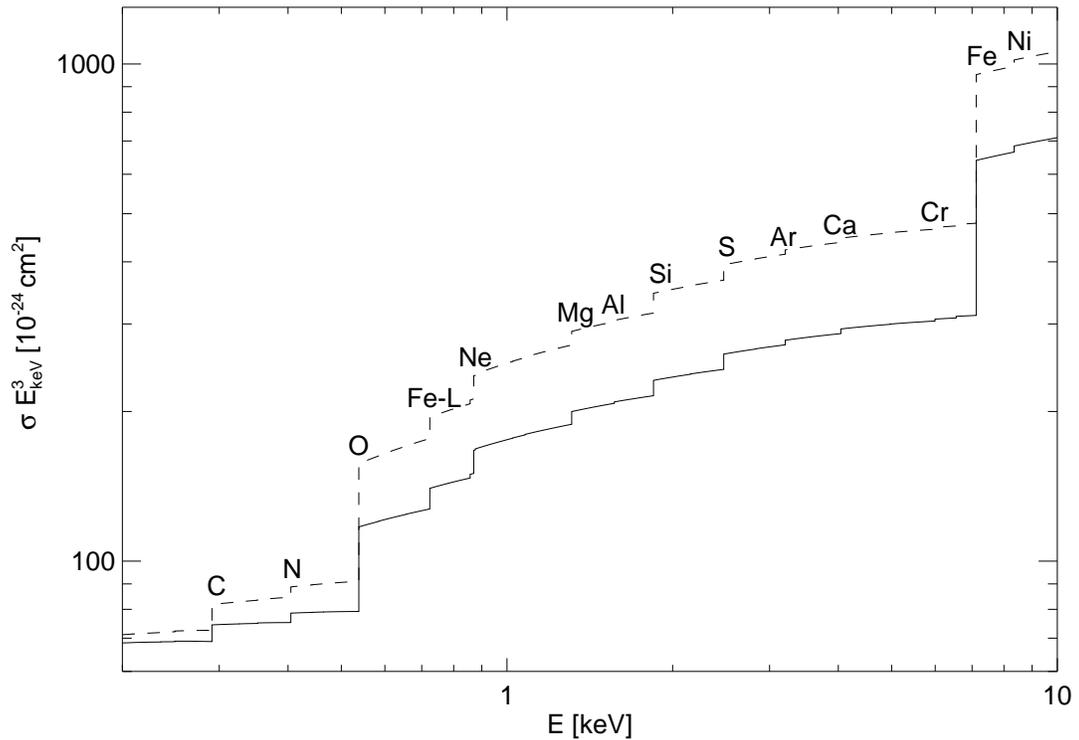}}
\caption{\protect\label{fig:solar}Absorptivity per hydrogen atom of the
  ISM for our adopted ISM abundances (solid line) and for the
  solar abundances of Tab.~\ref{tab:abund} (dashed line).  For
  clarity, the cross section has been multiplied with $E^3$;
  however, the $y$-axis is logarithmic here, not linear as in
  Fig.~\ref{fig:cosab}.  The revised abundances produce changes
  in $\sigism$ of up to 30\% with respect to the solar
  abundances. }
\end{figure*}

\begin{figure*}
\centerline{\includegraphics[width=0.8\textwidth]{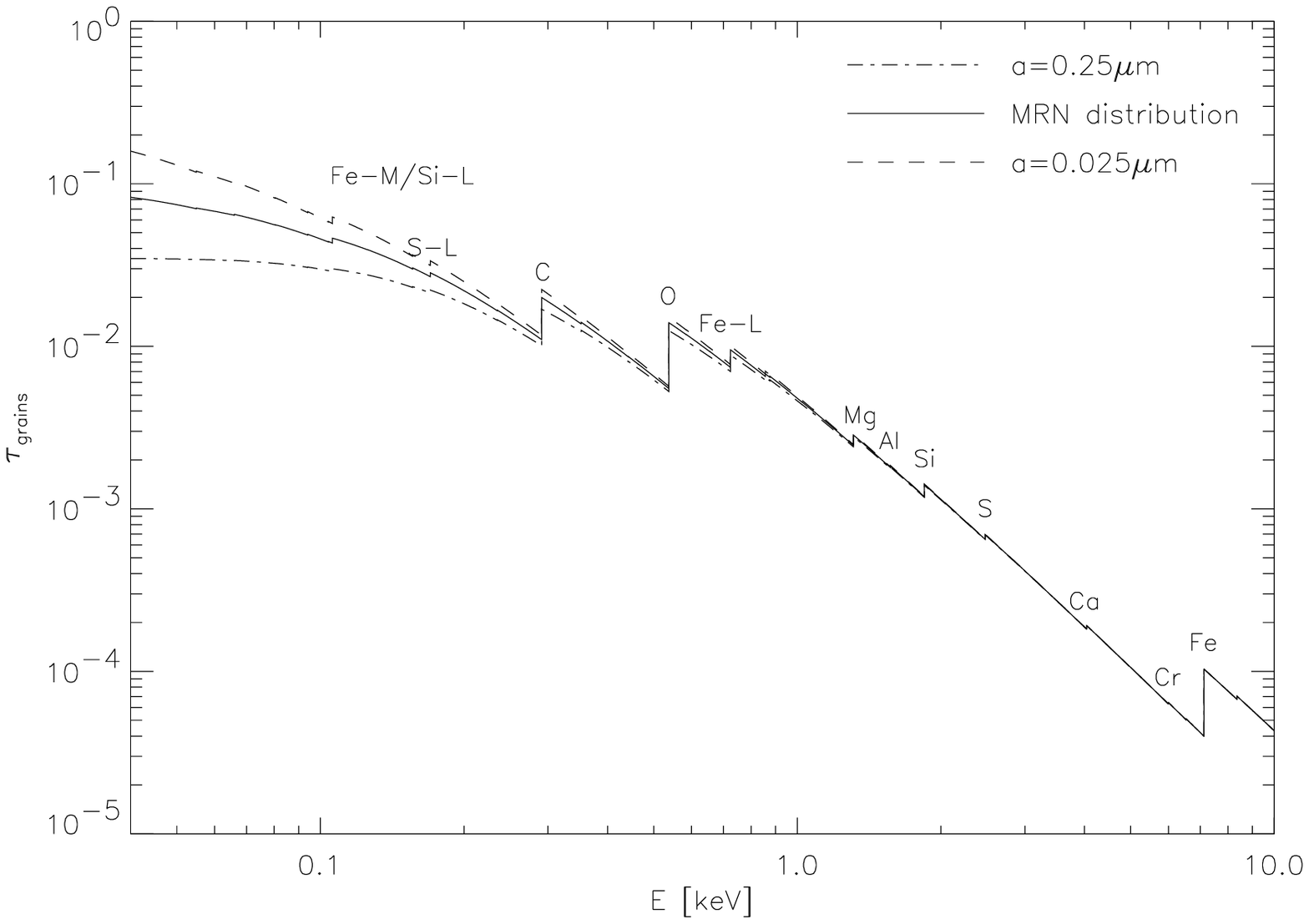}}
\caption{\protect\label{fig:graintau}Optical depth of grains for a hydrogen
  column $\nH=1.0 \times 10^{20} cm^{-2}$ (see Appendix~\ref{sec:f}).
  and for grains of radius $a=0.25\,\micron$ (dot dashed line) and
  $a=0.025\,\micron$ (dashed line) as a function of X-ray energy. The
  total mass in grains is constant. The shielding reduces the optical
  depth at low energies and for large grain sizes.  The labels denote
  the K-edges of the elements indicated.}
\end{figure*}

Also clear from Fig.~\ref{fig:cosab} is that the effect of grains on
$\sigism$ is small for a standard MRN distribution. The effect of
grains is found to be less than previous estimates such as MM83 in
part because our grain model consists of lower-density porous grains
and in part because we calculate $\sigma_{\rm grains}$ for an MRN
distribution of grain sizes rather than by choosing an average grain
size such as 0.3\,$\micron$ as used by \citeauthor{fireman:74a} and
others. Thus the greatest number of grains lies in very small grains
with $\tau\ll 1$ above $\sim 1$\,keV, so that self-shielding is not
important in these grains.  Consequently, for an MRN distribution
there are only small differences between the optical depth of an
entirely gas-phase ISM and the optical depth for an ISM in which some
of the gas atoms have been depleted into grains.  This is illustrated
in Fig.~\ref{fig:graintau}, where we plot the grain optical depth as a
function of energy for grains of radius $0.25\,\micron$ and
$0.025\,\micron$, as well as for an MRN distribution of grains.  These
optical depths are calculated using Appendix~\ref{sec:f} and
considering a hydrogen column density $\nH$ of $1.0 \times 10^{20}
{\rm cm}^{-2}$.  The total mass in grains is the same for all three
grain models shown.  Clearly, the self-blanketing factor affects the
opacity more for low energies and larger grain sizes.  Since the
change in grain optical depth for the MRN distribution is quite small,
grains only slightly modify the absorptivity of the ISM.

We stress, however, that the grain model as used here applies to the
diffuse ISM rather than to dense regions such as inner molecular
clouds or young stellar objects, where grains might play a much more
important role \citep{mathis:90a}. By changing the range of grain
sizes from the range adopted here for the diffuse ISM to a grain
distribution with larger grains, our formalism (including the XSPEC
routine described in Appendix~\ref{sec:xspec}) can be used to treat
these regions as well. As an example, we show in Fig.~\ref{fig:cosab}
the influence of grains with $a=0.3\,\micron$. In this case, X-ray
observations will allow the detection of the grains due to the large
influence of self-shielding, even with instruments that do not have
the high energy resolution to see the solid-state resonance effects
close to the absorption edges.

\subsection{Comparison with previous models}
As we outlined in Sect.~\ref{sec:atomic} and~\ref{sec:abundances}, the
major differences between this model and the previous models are the
updates to the atomic cross sections and the assumed abundances. With
the use of simulated observations, we can examine the impact of an
updated $\sigism$ on the interpretation of observations.

\begin{figure*}
\centerline{\includegraphics[width=0.8\textwidth]{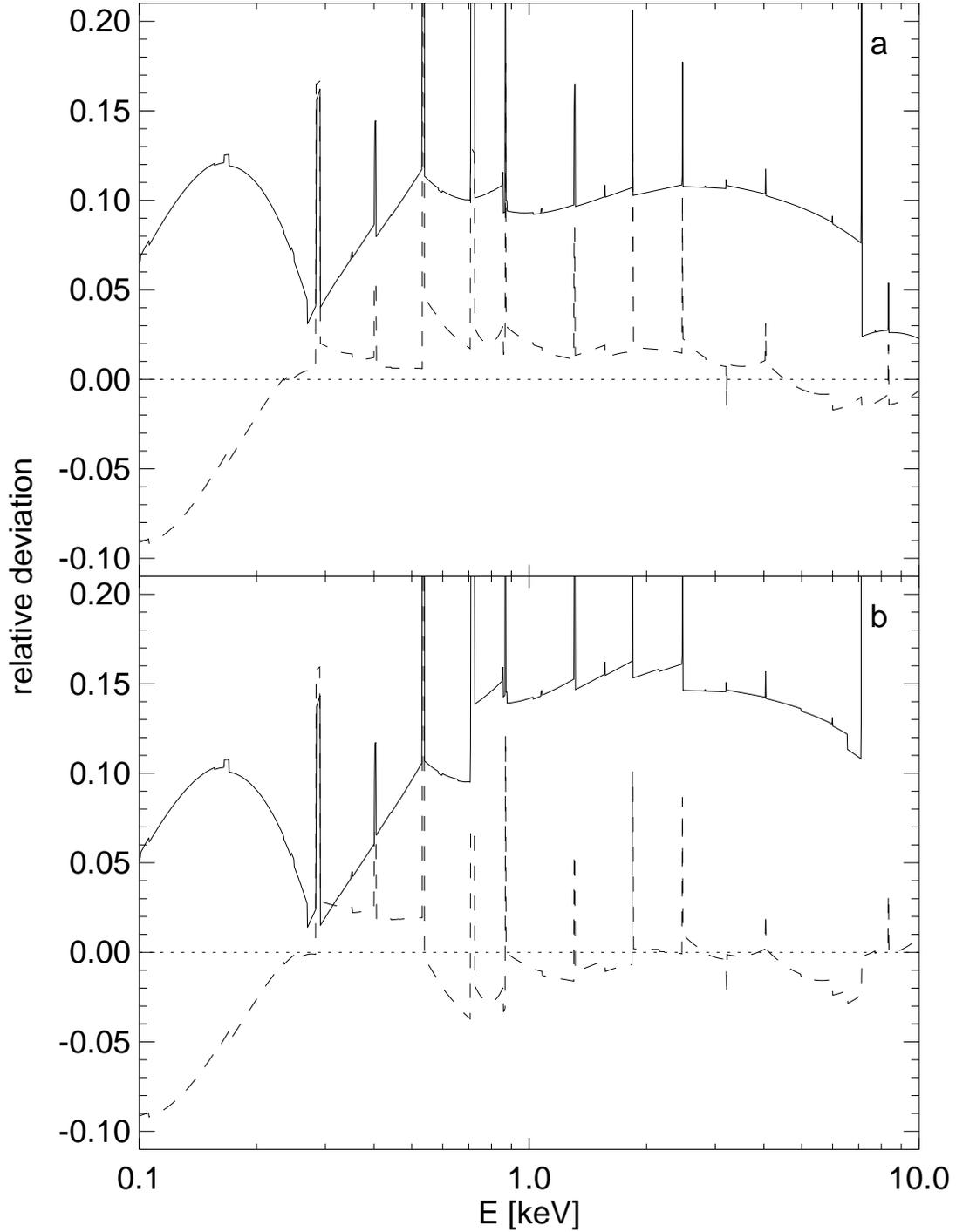}}
\caption{\protect\label{fig:coscmp}Relative deviation of the cross sections of
  BM92 (solid line) and MM83 (dashed) from the model described here.
  The relative computation is given as $(\sigma_i-\sigma)/\sigma$
  where $i=$BM92 or MM83, and where $\sigma$ is our formulation. The
  spikes at the K- and L-absorption edges are due to slight
  differences in the assumed edge energies. a: Relative deviation
  computed using the abundances of BM92 and MM83, respectively. This
  figure highlights the effect of the change in the cross-sections.
  b: Relative deviation to our model, computed using the solar
  abundances of Tab.~\ref{tab:abund}, illustrating the cumulative
  effect of using different default solar abundances and changing
  cross sections.  See \S\ref{sec:atomic} for further
  discussion.}
\end{figure*}

In order to demonstrate the effect of the updates to the cross
sections, we compare our values of $\sigism$ with those of MM83 and
BM92. In Fig.~\ref{fig:coscmp}a we show the relative deviation between
our formulation and that of the earlier models, using our adopted
cross-sections and the assumed abundances of the MM83 and BM92 models,
respectively. Thus, the figure shows the cumulative effect of the
updated cross sections. The differences between the models at photon
energies below 0.3\,keV are mainly a result of the uncertainty in the
He cross section. Additional peaks at the K- and L-shell edges are due
to $\sim 10$\,eV discrepancies in the assumed edge energy between
these models (\S\ref{sec:atomic}) and are not relevant for practical
work with moderate resolution instruments. The difference in the
values of $\sbf$ produces the curvature of the relative deviation. The
missing subshell absorption edges in the MM83 model are also readily
apparent.

In Fig.~\ref{fig:coscmp}b, we again show the relative deviation between our
model and that of the earlier models, but this time using our default
\emph{solar} abundance (Table~\ref{tab:abund}) for the abundances in our
model, including the additional elements not contained in the MM83 and BM92
models, as well as the H$_2$ molecule (important especially below $\sim
300$\,eV).  In this figure, all models are with ''solar" abundances, but
the disagreement between the ``solar'' abundances used by the different
authors results in 10\% of the difference between the three models compared
in Fig.~\ref{fig:coscmp}. However, the difference resulting from different
values of the ''solar" abundance is small compared to the difference that
results from using our default abundance values which are reduced with
respect to solar abundances.  The discrepancy in $\sigism$ which results
from the use of our default ISM abundances instead of the solar abundances
indicated in Table~\ref{tab:abund} is shown in Fig.~\ref{fig:solar}.

In order to illustrate the observable changes introduced by our
updated model of $\sigism$, we studied these effects using simulations
of observations. We simulated the observation of a bright X-ray source
which has a pure power-law spectrum with photon index $\Gamma=-1.7$,
and a normalization of $A_{\rm PL}=2\,\rm
ph\,\cm^{-2}\,s^{-1}\,keV^{-1}$ at 1\,keV. Such a source spectrum is
characteristic, e.g., of \objectname{Cygnus~X-1} in the hard state
\citep{dove:97c}.  We then used two values of hydrogen column,
$\nH=6\times 10^{21}\,\cm^{-2}$ and $\nH=6\times 10^{22}\,\cm^{-2}$,
to produce the absorbed spectrum using our model for $\sigism$, using
our assumed interstellar abundances.  The two values for $\nH$ are
necessary to illustrate the effects of $\nH$ both in the low energy
band (small $\nH$) and in the high energy band (large $\nH$).  We
simulated an observation by folding the photon spectrum through the
detector response matrix and adding Poisson noise. To ensure
comparable signal-to-noise ratios in all simulated observations, the
exposure times were chosen such that $10^{6}$ photons were contained
in the resulting spectrum. The detectors we chose to represent current
instruments were SIS on the Advanced Satellite for Cosmology and
Astronomy (ASCA) for the low $\nH$ simulation, and the Proportional
Counter Array (PCA) on the Rossi X-ray Timing Explorer (RXTE) for the
large $\nH$ simulation. The detector chosen to represent the quality
and energy resolution of the future missions is the European Photon
Imaging Camera (EPIC) pn CCD on XMM-Newton.  In neither case did we
take photon pileup or other instrumental effects such as a saturation
of the telemetry bandwidth into account.

\begin{figure*}
\centerline{\includegraphics[width=0.8\textwidth]{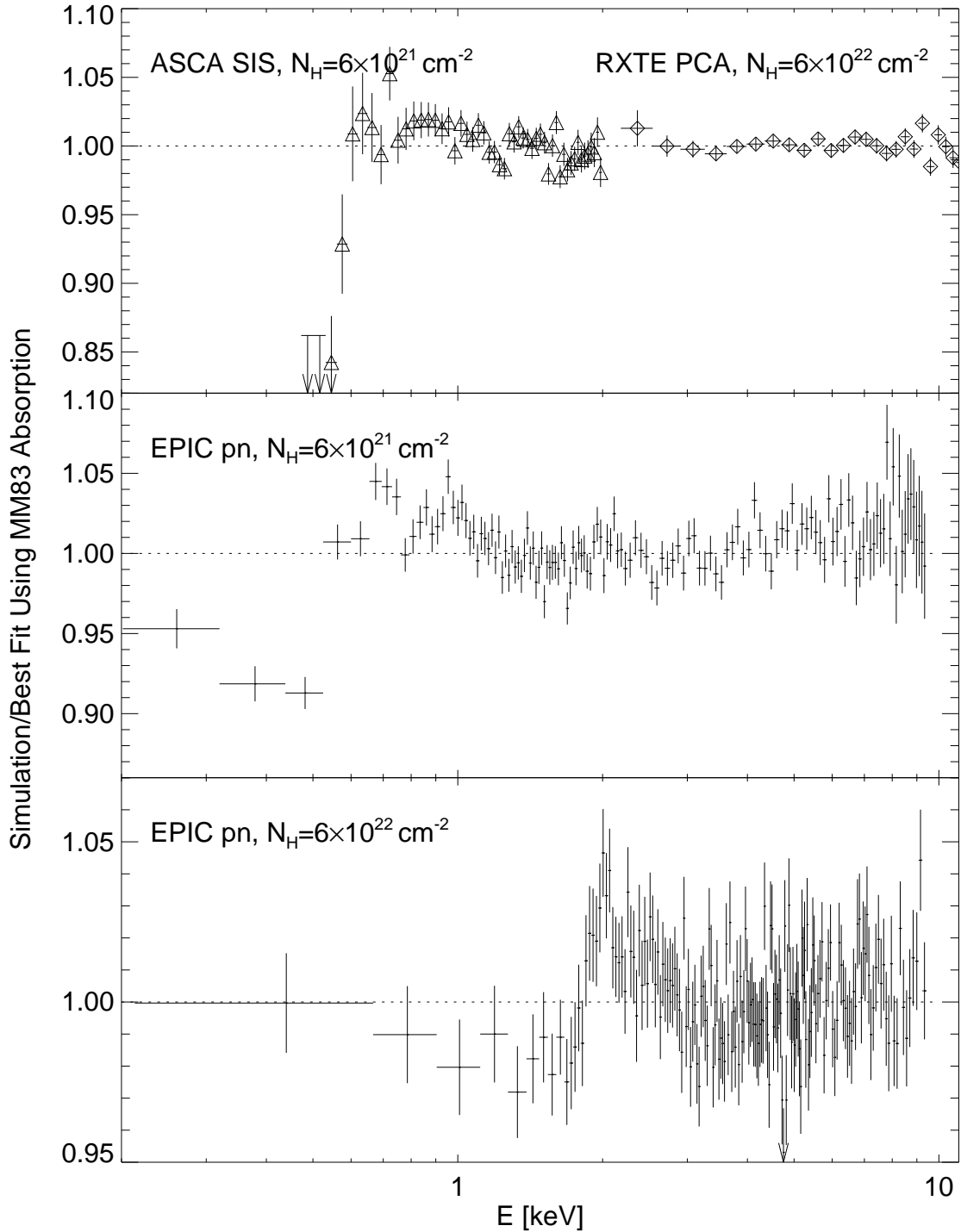}}
\caption{\protect\label{fig:simul}Ratio of simulated X-ray data absorbed
  with our model for $\sigism$ to the best fit model using a
  power-law model and the MM83 models, for the ASCA SIS
  (triangles), RXTE PCA (diamonds), and the XMM-Newton EPIC pn CCDs. The
  ASCA SIS simulation with our model is shown for energies below
  2\,keV.  In all cases, the error bars represent Poisson noise.
  While older models produce adequate fits for the current
  instruments (with the exception at the O edge, see text),
  differences of $\sim 5$\ldots 10\% are found with the EPIC pn
  simulations. }
\end{figure*}

The resulting simulated photon spectrum was rebinned and analyzed
using the MM83 and the BM92 models for $\sigism$. In
Fig.~\ref{fig:simul} we present the ratio between the data and the
resulting best-fit model using the MM83 model (the BM92 model without
varying the abundances gave qualitatively the same results). For the
PCA and the SIS, the fits with the MM83 model gave an acceptable
description of the data.  Our low $\nH$ simulation shows that the ASCA
SIS is able to detect the discrepancy in the oxygen abundance between
the solar and the ISM values. The other abundance differences are not
measurable in reasonable amounts of observing time. We note in passing
that \citet{ebisawa:96b} indeed found residuals in an observation of
the black hole candidate Cygnus X-1 that are very similar to those
shown in Fig.~\ref{fig:simul}, which points towards an oxygen
abundance of $\sim$80\% solar in the direction to Cyg~X-1.

For the large $\nH$ simulation, the fit to the simulated RXTE PCA
spectrum recovered the input parameters for $\Gamma$ and for
$A_{\rm PL}$ to two significant digits and found a hydrogen
column of $\nH=4\times 10^{22}\,\cm^{-2}$. The residuals shown in
Fig.~\ref{fig:simul} show some structure below 3\,keV that is
caused by underestimating $\nH$, which results in the non-optimal
$\chi^2$ value of the fit ($\chi^2_{\rm red}=1.7$). However, the
PCA has significant systematic uncertainties in this region
\citep{wilms:98c}, and the deviation seen in our simulation would
probably go undetected in a real observational situation.

Although we have not done an exhaustive search of the whole $\nH$
space, the above result convinces us that for most applications
using the current moderate resolution instruments, the models of
MM83 and BM92 are sufficient. The physical cause for the
underestimate of $\nH$ is that for energies above 1\,keV,
$\sigism$ is dominated by the metals. Since the abundance ratios
of the metals are a result of stellar nucleosynthesis, reducing
the metal abundance with respect to hydrogen from the MM83 or
BM92 values to our values can be compensated by assuming a
different (lower) hydrogen column.  Nor does the slightly
different shape of $\sigma_{\rm bf}$ that results from the
updated cross sections produce a noticeable effect in the PCA
simulations; the energy resolution of the PCA is insensitive to a
difference of this magnitude.

Our simulations for the EPIC pn camera, on the other hand, show
deviations on the 5\% level when fitted with the MM83 and BM92 models
(Fig.~\ref{fig:simul}). These changes are clearly due to the update of
the abundances and cross sections. It is for instruments such as the
EPIC pn camera that we recommend that our model be used. The EPIC
observation simulated here is quite short (the raw EPIC pn count-rate
of Cygnus~X-1 will be $\sim 2400$\,cps, so that the spectrum used for
modeling here has an exposure time of only 4\,ksec -- shorter than the
minimum XMM-Newton observing time of 5\,ksec), and the ratio plots in
Fig.~\ref{fig:simul} already show the characteristic curvature due to
the deviation in the shape of the cross section assumed in the MM83
models. We note that a similar behavior is also present when fitting
the simulated spectrum with the BM92 data.  When using the BM92 cross
sections, however, it is possible to obtain a good description of the
overall absorption by treating the abundances as free parameters.  In
this case the ``wavy'' residual structure still remains although the
relative abundances found for the elements do come close to the
abundances that we assumed in our modeling of $\sigism$. As a result,
we recommend using our model whenever high precision work on the
1\%--2\% level is necessary, as well as in cases when the absorbing
column resulting from the spectral fitting process is to be physically
interpreted (i.e., when comparing X-ray $\nH$ values with radio data,
or when modeling an observed X-ray spectrum for which the $\nH$ value
is known by other means).

\subsection{Caveats}
While we feel that the model for $\sigism$ as presented in the
previous sections is well-suited for most X-ray astronomical work,
several astronomical or physical issues had to be neglected in the
modeling process. In particular, our model does not include the
effects of the warm phase or of the ionized phase of the ISM.  In
addition, we ignore Thomson scattering, which becomes important at
moderate to high Thomson optical depths and $E\gtrsim 4$\,keV.  These
effects have not been included largely because they are negligible for
most situations in which our model is likely to be used.  Furthermore,
the model assumes that there is no variation of the abundances along
the line of sight. Finally, we note that energy-dependent
scattering by dust grains has not been considered here since an
XSPEC routine (\texttt{dust}) already exists to treat it.

The warm and ionized phases of the ISM have been neglected here
for a number of reasons. Large parts of the ISM are assumed to be
(moderately) ionized, and a good model for $\sigism$ should in
principle include the ionized phase. However, there is even less
agreement on the ionization state of the ISM as a whole than
there is on the grain phase, which was discussed in
\S\ref{sec:grain}.  The major reason for this disagreement is a
good one: the ionization of the ISM depends strongly on the line
of sight, so that any general model for the ionization structure
of the galaxy is prone to fail.  It appears, however, that most
of the warm phase of the ISM is only moderately ionized, so that
our model will still be applicable except for photon energies
close to the K-edges. For oxygen, e.g., the K-edge energy
increases from 0.54\,keV for \ion{O}{1} to 0.61\,\keV for
\ion{O}{4}, and the photoionization cross section is also
dependent on the ionization stage.  It should be noted, though,
that the ionization of hydrogen will strongly influence $\sigism$
for energies below 0.1\,keV. For instance, in their study of soft
X-ray emission from other galaxies, \citep{cui:96b}, assumed the
ratio between~\ion{H}{1} and~\ion{H}{2} to be~1.  Using the code
presented in Appendix~\ref{sec:xspec}, this effect may be taken
into account by changing the relative abundances of He and the
metals with respect to hydrogen.

Similar line of sight effects hold also true for the abundances. In
the galaxy, measurement of abundances in HII regions and planetary 
nebulae has suggested the existence of a radial abundance gradient,
particularly at small galactic radii \citep[and references therein]
{rana:91a,kaufer:94a}.  When computing the absorption of any background
object through such a medium with varying abundances, the abundances
measured in the X-rays will be the average of the abundance variations
along the line of sight. Since the abundances in our model are allowed to
vary, it is possible to take this effect into account. Obviously, in such a
case the assumed abundance set with respect to which the relative
abundances are measured is a matter of taste, and using the solar
abundances of Tab.~\ref{tab:abund} as a baseline instead of the ISM
abundances might be a better choice.

Our model also ignores the influence of Thomson scattering
because for many observations it will not make a substantial
contribution.  This is not to say that it is always negligible;
Thomson scattering is the dominant physical process for the
attenuation of X-rays above $\sim$10\,keV. As a rough guideline,
Thomson scattering should always be included in the modeling
process when $N_{\rm H} \gtrsim 10^{22}\,\rm cm^{-2}$ \citep[see,
e.g.,][]{stelzer:98a}.  Detailed instructions based on Monte
Carlo modeling of Thomson scattering in moderately optically
thick media are given by \citet{yaqoob:97a}. Note, however, that
\citeauthor{yaqoob:97a} assumes the electrons to be free. When
the electrons are bound, it is more appropriate to consider
incoherent scattering, i.e., Compton scattering, off the bound
electrons. The cross sections for incoherent scattering and the
Klein-Nishina cross section are slightly different (on the
percent level), which should be taken into account.  Analytical
estimates for this effect have been given by \citet{gorshkov:73a}
for the H atom; for other elements tables are available
\citep{hubbell:75a,henke:93a}. For most work, however, these
corrections are negligible.

Finally, we also stress that our model, like its predecessors MM83 and
BM92, is \emph{not} well suited for extremely high-precision work,
e.g., high signal-to-noise spectra with the grating spectrometers on
XMM-Newton and \textsl{Chandra}. The energy resolution of these
very-high-resolution instruments is so high that the X-ray absorption
fine structure (XAFS) within grains and molecules might become
relevant, as well as resonance effects close to the absorption edges
in the absorption cross section itself. For the inclusion of the
resonances much higher resolution cross sections than those available
here are required. XAFS effects become noticeable especially in
regions of large $\nH$ where grains are an important ingredient.  For
these studies, dedicated modeling codes are being devised, which
should be used instead.  See \citet{forrey:98a}, \citet{woo:97a},
\citet{woo:95a}, and the references therein, for further details.

\subsection{Summary and Outlook}
We have presented an improved model for the X-ray absorption in the
interstellar medium valid above $\sim 100$\,eV by updating the
relevant photoionization cross sections and by including the effect of
the H$_2$ molecule and a revised set of recommended elemental
abundances for the interstellar medium.  Since the abundances of the
metals are smaller than previously estimated, our values of $\sigism$
are generally smaller than those used previously. We have shown that
these updates will be relevant for instruments to be available as this
paper goes into print.  We have also reconsidered the effect of grains
in the ISM by improving the grain model and the calculation of
$\tau_{\rm grains}$, taking into account recent studies of the
physical properties of interstellar dust grains. Our results show that
the change in opacity due to grains on $\sigism$ for realistic grain
models and realistic chemical composition of the grains is small. As
described in Appendix~\ref{sec:xspec}, our calculation of $\sigism$ is
available as a subroutine that can be used with X-ray data analysis
packages such as XSPEC.

This work highlights the point made by \citet{shull:93a} and others,
that great uncertainties remain in much of the atomic physics data
needed to interpret X-ray astronomical observations.  The effective
areas and energy resolutions of instruments recently 
launched will be more sensitive than the available atomic data for the
X-ray energy range can accommodate, and the situation is even worse
for possible future instruments such as Constellation-X or XEUS.  The
differences between the currently available cross sections discussed
in \S\ref{sec:atomic} are still at the 5\% level.  We hope that the
advent of the new instruments will also initiate new attempts to more
accurately determine the atomic data on which X-ray astronomy can
rely. We will therefore continue over the next years to update the
database of the computations presented here and periodically release
new versions of the code presented in Appendix~\ref{sec:xspec} that
will incorporate the improvements in the abundances or the atomic
physics.

\acknowledgements This work has been partially financed by NASA Grants
NGT 5-80, NAG5-7339, and NAG5-7340, and by a National Physical Science
Consortium Fellowship grant to AA.  JW acknowledges the hospitality of
the high energy astrophysics groups at JILA and at the University of
California at San Diego where part of this work was done. We thank the 
referee Dan McCammon for his comments, which greatly improved the 
presentation of this paper, and for providing the code for the helium 
autionization resonances.  We also thank Laurie Kovalenko, Phil Maloney,
Katja Pottschmidt, Richard Rothschild, R\"udiger Staubert, Klaus Werner, 
and Svet Zhekov for useful comments.

\appendix

\section{Self-shielding in the grains}\label{sec:f}
To derive the optical depth of the grains (eq.~[\ref{eq:taudist}]) we
consider a population of grains of homogeneous chemical composition with
size $a$, geometrical cross section $\sigma_{\rm geom}$, column density
$\Nbar$, and average photoionization cross section $\sigbar$ (per atom).
For grains of realistic shape, $\Nbar$ is a complicated function of $a$.
Since the influence of grains on $\sigism$ is small, however, for
simplicity we can assume that the grains are spheres with radius $a$ so
that
\begin{equation}
\Nbar = \frac{4 \rho a}{3 \mu}=\frac{4na}{3}
\end{equation}
where $\rho$ is the mass density of the grain, $n$ the number
density of the grain, and $\mu$ the mean molecular weight of the
grains constituents. Using the notation of Section~\ref{sec:int},
$\mu$ is given by
\begin{equation}
 \mu  = \sum_Z A_Z \cdot \beta_Z \cdot \mu_Z \bigg/ \sum_Z
 A_Z\cdot \beta_Z
\end{equation}
Here, $\mu_Z$ is the molecular weight of element $Z$
(Tab.~\ref{tab:abund}). For the default abundances and depletions
of Tab.~\ref{tab:abund}, $\mu=18.2$\,amu\ $=3.0\times
10^{-23}$\,g/atom.  Assuming $\rho=1\,\rm g\,cm^{-3}$, the grain
number density is $n=3.3\times 10^{22}\,\rm atoms\,cm^{-3}$.  Ignoring
solid state effects, the photoionization cross section of the
grain material is given by
\begin{equation}
 \sigbar = \sum_Z A_Z \cdot \beta_Z \cdot \sbf(Z) \bigg/ \sum_Z
 A_Z\cdot \beta_Z
\end{equation}
For a column density $N_{\rm grains}$ of grains of a \emph{single size}
along the line of sight, the optical depth for photoabsorption of
the grains is the product between the geometrical optical depth
of the grains, $N_{\rm grains}\sigma_{\rm geom}$, and the fraction
of photons that encounter the grain and are absorbed:
\begin{equation}\label{eq:taugrain}
\tau_{\rm grains} = N_{\rm grains} \sigma_{\rm geom} 
                   \left( 1-\exp\left(-\sigbar \Nbar\right)\right).
\end{equation}
In computing $\sigism$ for grains of a single size it is
sometimes convenient to express $\tau_{\rm grains}$ in terms of
the optical depth that would be observed if the constituents of
the grain were in the gaseous phase, $\tau_{\rm grains} = f
N_{\rm gas} \sigbar$.  Hence,
\begin{equation}\label{eq:fonesize}
f=\frac{N_{\rm grains} \sigma_{\rm geom}}{N_{\rm gas} \sigbar}
                   \left( 1-\exp\left(-\sigbar \Nbar\right)\right)
= \frac{1-\exp\left(-\sigbar \Nbar\right)}{\sigbar\Nbar}.
\end{equation}
since the column density of the grain is $N_{\rm gas}/(N_{\rm
  grains}\sigma_{\rm geom})$.  This factor is called the
self-blanketing factor \citep{fireman:74a}. 

For the general case that the grains have a \emph{size distribution}
$\dd n_{\rm gr}(a)/\dd a$, where
\begin{equation}
\int_0^{\infty} \frac{\dd n_{\rm gr}}{\dd a} \dd a = 1
\end{equation}
a similar derivation shows that
\begin{equation}\label{eq:fallsize}
\tau_{\rm grains}=N_{\rm H} \xi_{\rm g}  
  \int_0^\infty \frac{\dd n_{\rm gr}(a)}{\dd a}  \sigma_{\rm geom} 
  \left(1-\exp\left(-\sigbar \Nbar\right)\right)
\dd a,
\end{equation}
where $\xi_{\rm g}$ is the number of grains per hydrogen atom along
the line of sight. For spherical grains,
\begin{equation}
\xi_{\rm g}= \sum A_Z\cdot\beta_Z\cdot\mu_Z \bigg/ 
  \int \frac{\dd n_{\rm gr}}{\dd a} \cdot \rho \cdot \frac{4}{3}
  \pi a^3 \cdot  \dd a 
\end{equation}
For a power-law size distribution, $\dd n_{\rm gr}(a)/\dd a = k
a^{-p}$, where $a_{\rm min} \le a \le a_{\rm amax}$, we can
express $\tau_{\rm grains}$ in terms of special functions.
Inserting $\sigma_{\rm geom}=\pi a^2$ into
equation~\eqref{eq:fallsize} and performing a partial integration, we
obtain for $p<4$
\begin{equation}
\begin{split}
\tau_{\rm grains}  &= N_{\rm H} \xi_{\rm g} \frac{\pi k}{p-3}
\Big\{
   \frac{1}{a_{\rm min}^{p-3}}
   \left( 1-{\rm e}^{-4\sigbar n a_{\rm min}/3}\right) 
   - 
   \frac{1}{a_{\rm max}^{p-3}}
   \left( 1-{\rm e}^{-4\sigbar n a_{\rm max}/3}\right) \\
 & \qquad +
     \left(4 \sigbar n \right/3)^{p-3}
     \left[
         \gamma\left(4-p, 4 \sigbar n a_{\rm max}/3\right) -
         \gamma\left(4-p, 4 \sigbar n a_{\rm min}/3\right)
     \right]
\Big\},
\end{split}
\end{equation}
where the incomplete gamma function is given by \citep[section
6.5]{abramowitz:64a}
\begin{equation}
\gamma(\alpha,x)=\int_0^x {\rm e}^{-t} t^{\alpha-1} \dd t
\end{equation}
and can be evaluated using algorithms presented by
\citet[Section~6.2]{nr}.

\section{Implementation into XSPEC}\label{sec:xspec}

Using the formalism and data described in Section~\ref{sec:model},
we have written subroutines that can be used in conjunction with
the popular X-ray data analysis package XSPEC \citep{arnaud:96a}.
These subroutines are available at:

\texttt{http://astro.uni-tuebingen.de/nh/}

Given the uncertainty of many of the parameters entering the
model, there are several interfaces to the subroutine. The main
spectral model, \texttt{tbabs}, is intended as a replacement for
the earlier \texttt{wabs} and \texttt{phabs} models, as well as
their redshifted versions. This model assumes the abundances and
depletion factors of Table~\ref{tab:abund}. A subroutine
\texttt{ztbabs} for modeling redshifted absorption is also
available. The latter subroutine does not account for the dust
component, as X-ray observations with a measurable dust influence
will most probably only include observations of galactic objects,
at least within the next decade.

In addition, the XSPEC model \texttt{tbvarabs} provides a full
interface to the X-ray absorption model presented in
Section~\ref{sec:model} and allows the user to change all
abundances, depletion factors, and grain properties. We stress
that this model is very powerful and therefore prone to misuse.
Caution is required when fitting data with too many free
parameters.

\begin{deluxetable}{llllll}
\tablecaption{Fit coefficients $a_i$ to equation~\eqref{eq:h2fit} for the
  H$_2$ photoabsorption cross section.\label{tab:h2tab}}
\tablecolumns{6}
\tablewidth{0pt}
\tablehead{
\colhead{$a_0$} & \colhead{$a_1$}   & \colhead{$a_2$}  &  \colhead{$a_3$}    &  \colhead{$a_4$}  & \colhead{$a_5$}}
\startdata
0.664 & -11.768 & 78.118 & -231.339 & 368.053 & -189.953\\
\enddata
\end{deluxetable}

\begin{deluxetable}{rlcccccccc}

\tablecaption{Molecular weight, K edge energies,
  abundances and depletion factors $1-\beta_Z$ for the abundant
  elements.\protect\label{tab:abund} } 
\tablecolumns{8}
\tablewidth{0pt}
\tablehead{
\multicolumn{2}{c}{Element} &
\colhead{$\mu_Z$\tablenotemark{a}} &
\colhead{$E_{\rm K}$\tablenotemark{b}} &
\multicolumn{2}{c}{$12+\log A_Z$} &
\multicolumn{3}{c}{$1-\beta_Z$\tablenotemark{e}} \\
& & & 
\colhead{keV} &
\colhead{solar\tablenotemark{c}} & 
\colhead{ISM\tablenotemark{d}} & 
\colhead{this paper} &
\colhead{MM83\tablenotemark{f}} &
\colhead{Ride77\tablenotemark{g}}}
\startdata
 1 & H  &\phn{}1 &\nodata &      12.00 &     12.00 & 1.0\phm{00}& 1.0\phn & 1.0 \\
 2 & He &\phn{}4 &\nodata &      10.99 &     10.99 & 1.0\phm{00}& 1.0\phn & 1.0 \\
 6 & C  & 12     & 0.29   & \phn{}8.60 &\phn{}8.38 & 0.5\phm{00}& 0.0\phn & 0.2 \\
 7 & N  & 14     & 0.41   & \phn{}7.97 &\phn{}7.88 & 1.0\phm{00}& 0.0\phn & 0.5 \\
 8 & O  & 16     & 0.54   & \phn{}8.93 &\phn{}8.69 & 0.6\phm{00}& 0.75    & 0.5 \\
10 & Ne & 20     & 0.87   & \phn{}8.09 &\phn{}7.94 & 1.0\phm{00}& 1.0\phn & 1.0 \\
11 & Na & 23     & 1.08   & \phn{}6.31 &\phn{}6.16 & 0.25\phn   & 0.0\phn &\nodata \\
12 & Mg & 24     & 1.31   & \phn{}7.59 &\phn{}7.40 & 0.2\phm{00}& 0.0\phn & 0.2 \\
13 & Al & 27     & 1.57   & \phn{}6.48 &\phn{}6.33 & 0.02\phn   & 0.0\phn &\nodata \\
14 & Si & 28     & 1.85   & \phn{}7.55 &\phn{}7.27 & 0.1\phm{00}& 0.0\phn & 0.5 \\
15 & P  & 31     & 2.15   & \phn{}5.57 &\phn{}5.42 & 0.6\phm{00}& \nodata &\nodata \\
16 & S  & 32     & 2.48   & \phn{}7.27 &\phn{}7.09 & 0.6\phm{00}& 0.0\phn & 0.7 \\
17 & Cl & 35     & 2.83   & \phn{}5.27 &\phn{}5.12 & 0.5\phm{00}& 0.0\phn &\nodata \\
18 & Ar & 40     & 3.20   & \phn{}6.56 &\phn{}6.41 & 1.0\phm{00}& 1.0\phn & 0.5 \\
20 & Ca & 40     & 4.04   & \phn{}6.34 &\phn{}6.20 & 0.003      & 0.0\phn &\nodata \\
22 & Ti & 48     & 4.97   & \phn{}4.93 &\phn{}4.81 & 0.002      & \nodata &\nodata \\
24 & Cr & 52     & 5.97   & \phn{}5.68 &\phn{}5.51 & 0.03\phn   & 0.0\phn &\nodata \\
25 & Mn & 55     & 6.55   & \phn{}5.53 &\phn{}5.34 & 0.07\phn   & \nodata &\nodata \\
26 & Fe & 56     & 7.12   & \phn{}7.50 &\phn{}7.43 & 0.3\phm{00}& 0.0\phn & 0.2    \\
27 & Co & 59     & 7.73   & \phn{}4.92 &\phn{}4.92 & 0.05\phn   & \nodata &\nodata \\
28 & Ni & 59     & 8.35   & \phn{}6.25 &\phn{}6.05 & 0.04\phn   & 0.0\phn &\nodata \\
\enddata
\tablenotetext{a}{molecular weight in amu,}
\tablenotetext{b}{K edge energy \citep[rounded after][]{verner:95a}}
\tablenotetext{c}{solar abundance \citep[see
also \S\ref{sec:abundances}]{anders:89a,grevesse:91b,grevesse:93a},}
\tablenotetext{d}{adopted abundance of the ISM based on \citet{snow:96a},
\citet{cardelli:96a}, \citet{meyer:97a} and \citet{meyer:98a}, 
see \S\ref{sec:abundances};}
\tablenotetext{e}{ratio of gas abundance to total ISM abundance,
  $1-\beta_Z$, 
using our adopted abundances (\S\ref{sec:grain}). For comparison, the
older values of}
\tablenotetext{f}{\citet{morrison:83a}, and}
\tablenotetext{g}{\citet{ride:77a} are also listed.}
\end{deluxetable}

\end{document}